\documentclass[aip,longbiblography
]{revtex4-1}
\usepackage{chemformula}
\usepackage{siunitx}
\usepackage{pdfpages}
\makeatletter
\AtBeginDocument{\let\LS@rot\@undefined}
\makeatother
\usepackage{graphicx}
\usepackage{amsmath,amsfonts,amssymb}
\usepackage[section]{placeins}
\usepackage{xcolor}
\definecolor{betul}{cmyk}{0.8, 0.2, 0.8, 0.0} 

\usepackage{xcolor}
\definecolor{Tobias}{cmyk}{0.0, 0.7, 0.7, 0.0} 

\draft 
\usepackage{mathrsfs}
\usepackage[scr=rsfs,cal=boondox]{mathalfa}
\begin{document}


\title[Above Room Temperature Ferroelectricity in Epitaxially Strained \ch{KTaO3}]{Above Room Temperature Ferroelectricity in Epitaxially Strained \ch{KTaO3}} 



\author{Tobias Schwaigert}
\thanks{Authors contributed equally}
\affiliation{Platform for the Accelerated Realization, Analysis, and Discovery of Interface Materials (PARADIM), Cornell University, Ithaca, New York 14853, USA}
\affiliation{Department of Materials Science and Engineering, Cornell University, Ithaca, NY 14853, USA}
\author{Salva Salmani-Rezaie}
\thanks{Authors contributed equally}
\affiliation{School of Applied and Engineering Physics, Cornell University, Ithaca, New York 14853, USA}
\affiliation{Kavli Institute at Cornell for Nanoscale Science, Cornell University, Ithaca, New York 14853, USA}
\author{Sankalpa Hazra}
\thanks{Authors contributed equally}
\affiliation{Material Science and Engineering, Pennsylvania State University, University Park, PA 16802, USA}
\author{Utkarsh Saha}
\affiliation{Material Science and Engineering, Pennsylvania State University, University Park, PA 16802, USA}
\author{Maya Ramesh}
\affiliation{Department of Materials Science and Engineering, Cornell University, Ithaca, NY 14853, USA}
\author{Aiden Ross}
\affiliation{Material Science and Engineering, Pennsylvania State University, University Park, PA 16802, USA}
\author{Bet\"{u}l Pamuk}
\affiliation{Department of Physics and Astronomy, Williams College, Williamstown, Massachusetts 01267, USA}
\author{Long-Qing Chen}
\affiliation{Material Science and Engineering, Pennsylvania State University, University Park, PA 16802, USA}
\author{David A. Muller}
\affiliation{School of Applied and Engineering Physics, Cornell University, Ithaca, New York 14853, USA}
\affiliation{Kavli Institute at Cornell for Nanoscale Science, Cornell University, Ithaca, New York 14853, USA}
\author{Darrell G. Schlom}
\affiliation{Platform for the Accelerated Realization, Analysis, and Discovery of Interface Materials (PARADIM), Cornell University, Ithaca, New York 14853, USA}
\affiliation{Department of Materials Science and Engineering, Cornell University, Ithaca, NY 14853, USA}
\affiliation{Kavli Institute at Cornell for Nanoscale Science, Cornell University, Ithaca, New York 14853, USA}
\affiliation{Leibniz-Institut für Kristallzüchtung, Max-Born-Str. 2, 12489 Berlin, Germany}
\author{Venkatraman Gopalan}
\affiliation{Material Science and Engineering, Pennsylvania State University, University Park, PA 16802, USA}
\affiliation{Deparment of Physics, Pennsylvania State University, University Park, PA 168010, USA}
\author{Kaveh Ahadi}
\email[ahadi.4@osu.edu]{}
\affiliation{Department of Materials Science and Engineering, Ohio State University, Columbus, OH 43210 USA}
\affiliation{Department of Electrical and Computer Engineering, Ohio State University, Columbus, OH 43210 USA}

\date{\today }

\begin{abstract}
 Epitaxial strain is a powerful means to engineer emergent phenomena in thin films and heterostructures. Here, we demonstrate that \ch{KTaO3}, a cubic perovskite in bulk form, can be epitaxially strained into a highly tunable ferroelectric. \ch{KTaO3} films grown commensurate to \ch{SrTiO3} (001) substrates experience an in-plane strain of -2.1 \% that transforms the cubic structure into a tetragonal polar phase with transition temperature of \SI{475}{K}, consistent with our thermodynamic calculations. We show that the Curie temperature and the spontaneous electric polarization can be systematically controlled with epitaxial strain. Scanning transmission electron microscopy reveals cooperative polar displacements of the potassium columns with respect to the neighboring tantalum columns at room temperature. Optical second-harmonic generation results are described by a tetragonal polar point group ($4mm$), indicating the emergence of a global polar ground state. We observe a ferroelectric hysteresis response, using metal-insulator-metal capacitor test structures. The results demonstrate a robust intrinsic ferroelectric state in epitaxially strained \ch{KTaO3} thin films.
 
\end{abstract}


\maketitle 

\section{Introduction\label{sec:level1}}
The ability to precisely tune specific properties defines the success of any materials system as a platform for new technologies. The misfit to an underlying substrate can impart an enormous strain to films and heterostructures, exceeding the failure limit of thick bulk samples, and modifying their energy landscape. Consequently, epitaxial strain can be harnessed as a tuning knob for emergent phenomena and functional properties. For example, epitaxial strain enhancement of transistor mobility\cite{nguyen1992} or the transition temperature in ferromagnets,\cite{beach1993,fuchs2008} ferroelectrics,\cite{choi2004} and superconductors\cite{bozovic2002} have been achieved. Even more exciting is when the epitaxial strain reveals a hidden ground state, such as ferromagnetism,\cite{lee2010strong} ferroelectricity,\cite{haeni2004, lee2010strong} and superconductivity,\cite{ruf2021strain} that are absent in the unstrained bulk form of that material system.

Thin films of ferroelectric materials play an essential role in high-density memory devices,\cite{scott1989} which makes the ability to control their ferroelectricity of great interest. For practical device applications and fundamental studies, it is equally desirable to tune the ferroelectric transition over a wide temperature range, from above room temperature to cryogenic temperatures, enabling operation in diverse environments and revealing temperature-dependent phenomena. Strong coupling between strain and ferroelectricity makes epitaxial strain a particularly efficient tuning knob to control electric polarization and ferroelectricity. \cite{devonshire1954theory, pertsev1998effect}

The efficiency of strain control of ferroelectricity is captured by the strain derivatives of key properties, such as $dT_c/d\varepsilon$\% and $dP_r/d\varepsilon$\%, which quantify the change in transition temperature and remanent polarization per unit strain. For example, while the bulk \ch{SrTiO3} is paraelectric, epitaxially strained \ch{SrTiO3} films exhibit emergent ferroelectricity with Curie temperatures reaching room temperature.\cite{haeni2004,biegalski_epitaxialy_2006} $dT_c/d\varepsilon$ and $dP_r/d\varepsilon$ are \SI{83}{K} and \SI{9}{\micro\coulomb\per\square\centi\metre}, respectively, for \ch{SrTiO3}.\cite{sheng2010modified} \ch{KTaO3} is a quantum paraelectric, that is close to a polar instability but remains paraelectric down to low temperature. With $dT_c/d\varepsilon$ = \SI{361}{K} and  $dP_r/d\varepsilon$ = \SI{9}{\micro\coulomb\per\square\centi\metre} \ch{KTaO3} is an ideal candidate for epitaxial strain tuning of ferroelectricity despite its cubic ground state. \cite{uwe1975electrostriction, hazra2024colossal}

Tunable electric polarization coupled with enhanced spin-orbit coupling, compared to \ch{SrTiO3},\cite{nakamura2009} make \ch{KTaO3} an intriguing candidate for spintronics applications.\cite{gupta2022} The discovery of a 2D electron gas (2DEG)\cite{zou2015} and superconductivity\cite{liu2021, ueno2011discovery} at \ch{KTaO3} interfaces combined with the reports of efficient spin-to-charge interconversion\cite{vicente2021spin, al2025spin} mark \ch{KTaO3} as an ideal model system to explore these phenomena in the presence of switchable spontanous polarization. Furthermore, theoretical investigations suggest that Rashba spin splitting can be efficiently improved with the application of compressive stress.\cite{wu2020strain}

\ch{KTaO3} is primarily studied in bulk due to the lack of high-quality epitaxial thin films, which are critical for tuning its physical properties and any potential device applications. High-quality films grown under adsoprtion-controlled conditions by molecular-beam epitaxy (MBE)\cite{schwaigert2023} and hybrid pulsed-laser depostion\cite{kim2024electronic}  have recently become available, enabling epitaxial strain engineering of \ch{KTaO3} and experimental testing of theoretical predictions.\cite{tao2016strain,tyunina2010} Here, we demonstrate the epitaxial stabilization of the ferroelectric tetragonal polymorph of \ch{KTaO3} by imposing epitaxial compressive strain through commensurate epitaxial growth on various lattice mismatched substrates under in-plane compressive strain. The emergence of ferroelectricity is supported by DFT and thermodynamic calculations, the atomic-scale characterization of the polar displacements using HAADF-STEM, second harmonic generation (SHG) rotational anisotropy response suggesting a $4mm$ point group, and the observation of ferroelectric hysteresis in metal-insulator-metal (MIM) hetero-structures.

\section{Results}
\subsection*{Theoretical considerations on the formation of a polar point group}

\begin{figure*}[h]
\includegraphics[width=\textwidth]{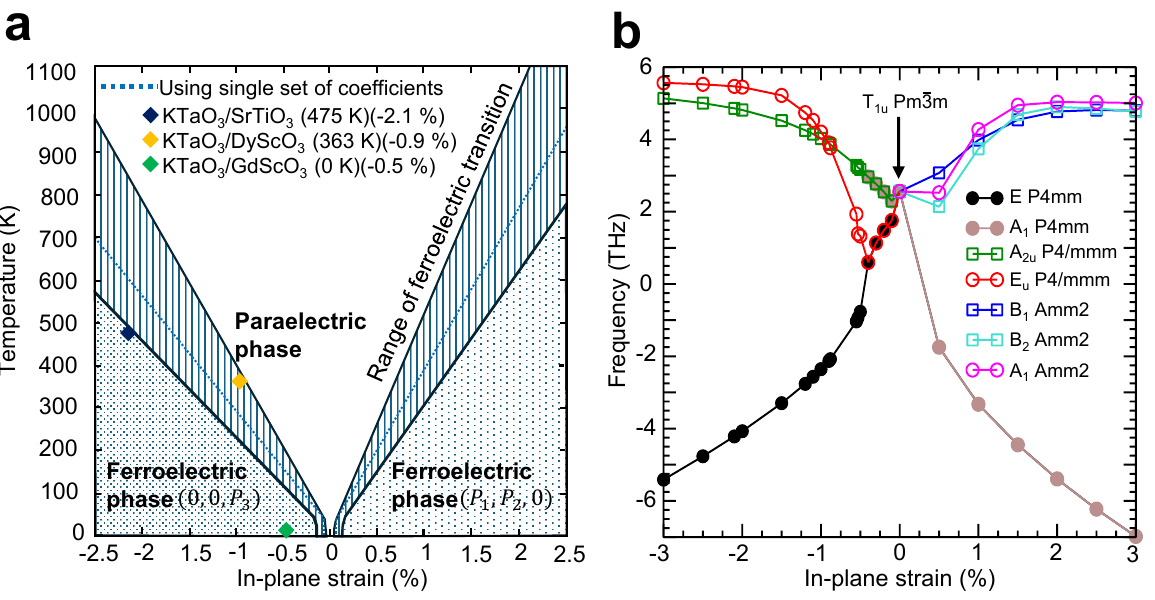}
\caption{\label{fig:phasefield} \textbf{Thermodynamic and DFT investigation of the epitaxial strain control of ferroelectricity in \ch{KTaO3}}. \textbf{a} Expected shift in $T_c$ of (001) \ch{KTaO3}, under epitaxial strain. The experimental values of $T_c$  are from second-harmonic generation (SHG) experiments (Fig. 4 and Fig. S11). The components of the spontaneous polarization are given with respect to the pseudocubic axes of \ch{KTaO3}.  \textbf{b} Lowest phonon frequencies of the structures with the $P4/mmm$, $P4mm$, and $Amm2$ symmetries as a function of epitaxial strain, with imaginary phonon frequencies presented as negative. Tetragonal structure with the P4/mmm symmetry starts to have imaginary phonon frequency beyond 0.5\% compressive (tensile) strain with the $A_{2u}$ ($E_u$) symmetry. Freezing in this phonon mode leads to a ferroelectric phase transition to a structure with the $P4mm$ ($Amm2$) symmetry.}
\end{figure*}

\ch{KTaO3} is a cubic perovskite with a lattice constant of a$_{KTO}$ = 3.988 \AA\cite{zhurova2000} at room temperature. The theoretical lattice parameter of cubic perovskite \ch{KTaO3} with $Pm\overline{3}m$ ($\#221$) symmetry is 3.990 \AA, resolved from first-principles calculations. As the structure is compressively strained to simulate the properties when grown on \ch{GdScO3} (-0.5\%), \ch{DyScO3} (-0.9\%), and \ch{SrTiO3} (-2.1\%), the out-of-plane lattice parameter is relaxed for each tetragonal structure with the $P4/mmm$ (space group $\#123$) symmetry.
The ferroelectric transition temperature can be calculated by solving the equation $T_c$ = max($T^{1}_{C}$,$T^{2}_{C}$) where $T^{1}_{C}$ and $T^{2}_{C}$ are solutions to\cite{li2006thermo}
\begin{equation}
    \alpha_{1}\left(T^{1}_{C} \right) + \Delta\alpha_{1} = \alpha_{1}T^{1}_{C} - \frac{\left(Q_{11}+Q_{12}\right)\epsilon_s}{s_{11} + s_{12}} = 0
\end{equation}
and
\begin{equation}
    \alpha_{1}\left(T^{2}_{C} \right) + \Delta\alpha_{3} = \alpha_{1}T^{2}_{C} - \frac{2Q_{12}\epsilon_s}{s_{11} + s_{12}} = 0
\end{equation}
where $\alpha_{1}$ represents the first order Landau coefficient, $\Delta\alpha_{1}$ and $\Delta\alpha_{3}$  denote the changes to the $\alpha_{1}$ coefficient due to strain along the enplane and out of plane directions respectively, $Q_{11}$ and $Q_{12}$ are components of the electrostrictive tensor and  $s_{11}$, $s_{12}$ are the components of the elastic compliance tensor, $\epsilon_{s}$ is the misfit strain and $T^{1}_{C}$ and $T^{2}_{C}$ are the transition temperatures due to the compressive and tensile strain respectively. This allows us to construct the phase transition boundaries between the paraelectric and ferroelectric phases in the temperature-strain phase diagram marked by the dotted lines as illustrated in Fig. \ref{fig:phasefield}. The dotted boundaries are calculated using a single set of coefficients (Table S2-3). An extended derivation can be found in the Supporting Information II.

The paraelectric ($4/mmm$) phase in the figure has zero net polarization, i.e., $P_1 = P_2 =  P_3 =0$. The ferroelectric phases are the tetragonal ($4mm$) phase with $P_1 = P_2 = 0$,  $P_3 \neq 0$ and out-of-plane polarization, induced by a compressive in-plane stress, and the orthorhombic ($mm2$) phase  $P_1 = P_2 = 0$, $P_3 \neq 0 $ showing in-plane polarization due to the tensile in-plane stress.

Phonon calculations of the tetragonal structures with the $P4/mmm$ symmetry reveal a $\Gamma$-point instability with the $A_{2u}$ symmetry and $\Gamma_3^- (a)$ irreducible representation, with eigenfrequency of $0.94i$, $2.57i$, and $4.22i$ THz, for each compressive strain, respectively. When the displacements of the atoms along this phonon frequency with $\omega=4.22i$ are frozen into the tetragonal atomic structure, the symmetry is reduced to $P4mm$ (\#99). This space group is energetically favored by 14 meV per formula unit compared to the parent $P4/mmm$ structure, for growth on SrTiO$_3$ with -2.1\% compressive strain. The out-of-plane phonon frequency shows a significant hardening from the $A_{2u}$ mode of the $P4/mmm$ structure to the $A_1$ mode of the $P4mm$ structure, indicating a second-order phase transition to the ferroelectric $P4mm$ structure upon compressive strain. The in-plane phonon frequency with the $E$ symmetry is not significantly affected with increasing compressive strain. Upon application of tensile strain, $P4/mmm$ symmetry structure has a $\Gamma$-point instability with the $E_{u}$ symmetry and $\Gamma_5^- (a,-a)$ irreducible representation. At 0.5\% tensile strain, this phonon mode is already hardened with a phase transition to the tetragonal ferroelectric structure with the $Amm2$ symmetry and lowest phonon frequencies of $A_1$, $B_1$ and $B_2$ are presented in Fig 1. The full band diagram of unstrained KTO (Fig. S1) and under -2.1 \% strain (Fig. S2) can be found in the Supporting Information Section I. The exact strain value that this transition occurs strongly depends on the exchange and correlational functional choice of DFT. Here, we choose the PBEsol functional because it is stable at the $Pm\bar{3}m$ structure, in agreement with the literature, \cite{Spaldin2023} allowing us to study the effects of strain separately, while the SCAN functional is not stable and quantum anharmonic effects play a significant role in KTaO$_3$, which is beyond the scope of this manuscript. \cite{Franchini2023, Franchini2024}


\subsection*{Growth and Structural Characterization}
\begin{figure*}[b]
\includegraphics[width=\textwidth]{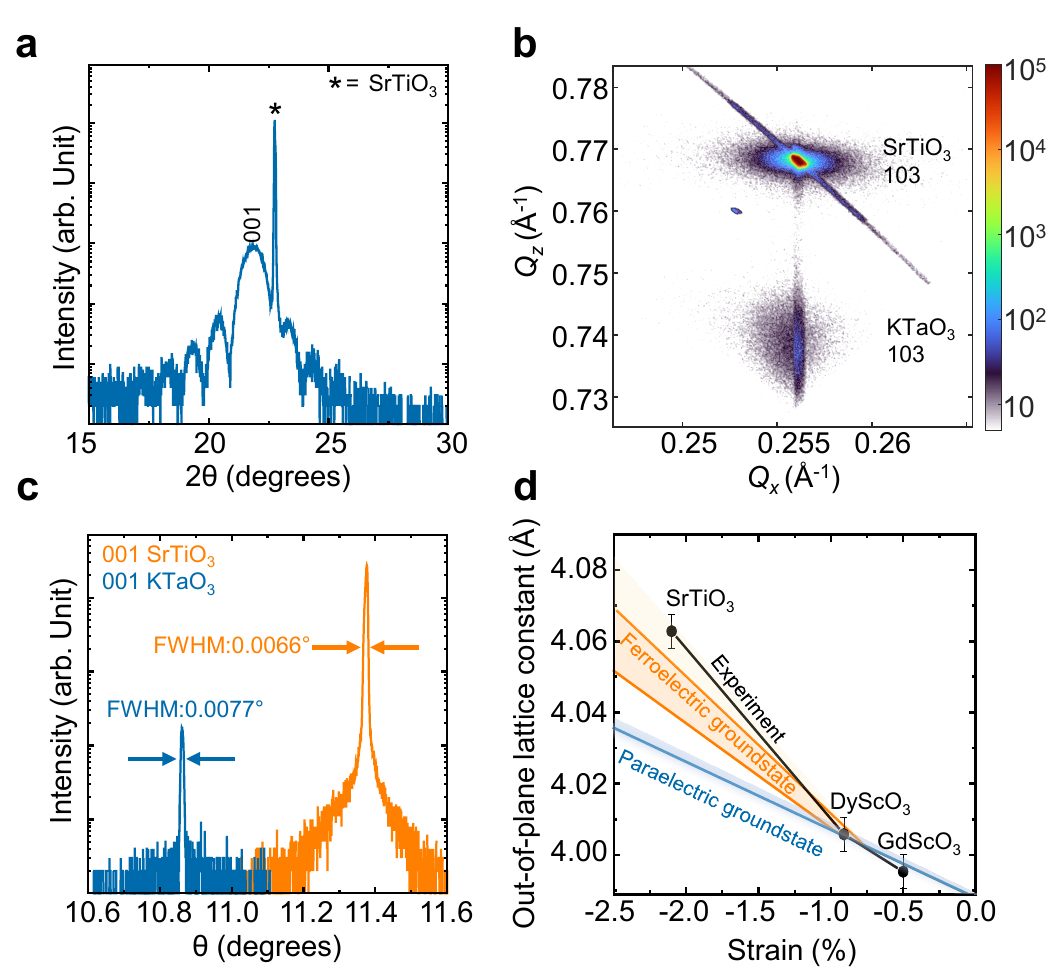}
\caption{\label{fig:XRD} \textbf{X-Ray diffraction of a 9.3 nm thick film of \ch{KTaO3} grown on \ch{SrTiO3} (001).} \textbf{a} $\theta$-2$\theta$ scan showing the 001 reflection of \ch{KTaO3} and \ch{SrTiO3}. Symmetric Laue fringes indicate a well-defined film thickness, indicative of an abrupt interface between film and substrate (asterisks * denotes substrate reflections). \textbf{b} X-ray reciprocal space mapping of the same 9.3 nm thick \ch{KTaO3} film, confirming that the film is commensurately strained to the  \ch{SrTiO3} substrate. \textbf{c} Overlaid rocking curves of the 001 \ch{KTaO3} and \ch{SrTiO3} peaks, showing comparable FWHMs, indicating low out-of-plane mosaicity ($\Delta\omega \approx$ 0.008°). \textbf{d} Measured out-of-plane lattice constant of \ch{KTaO3} thin films on various substrates against the expected lattice constants from thermodynamic analysis of the paraelectric (blue) and the ferroelectric (orange) ground states. The range in the calculated out-of-plane lattice constant is due to the range in the reported sets of coefficients (Table S2-3). }
\end{figure*}

\ch{KTaO3} grows ``cube-on-cube" on \ch{SrTiO3} (001) substrates and ``cube-on-pseudocube" on \ch{DyScO3} (110)$_o$ (pseudo-cubic lattice-constant 3.947 \AA\cite{uecker2008}, where the subscript $O$ indicates orthorhombic indices) and \ch{GdScO3} (110)$_o$ (pseudo-cubic lattice-constant 3.967\AA\cite{uecker2008}). The MBE growth details are described elsewhere.\cite{schwaigert2023} Here, all the films are grown in an adsorption-controlled regime with a K:Ta excess flux ratio of about 10:1. X-ray diffraction (XRD) was used to characterize the structural quality of grown films. XRD of the \ch{KTaO3}/\ch{SrTiO3} film is shown in Fig. \ref{fig:XRD}a; the \ch{KTaO3}/\ch{DyScO3} and \ch{KTaO3}/\ch{GdScO3} samples are shown in the Supportingl Information (Fig. S3). The $\theta$-2$\theta$ scan shows only 001 peaks, confirming that the film is single-phase and oriented with its \textit{c}-axis perpendicular to the plane of the substrate.

X-ray reciprocal space mapping (RSM) around \ch{SrTiO3} and \ch{KTaO3} 103 reflections confirm that the films are coherently strained to the substrate (Fig. \ref{fig:XRD}b). The full width at half maximum (FWHM) of the rocking cruve of the  \ch{KTaO3} ($\Delta\omega\approx$ 0.008$^{\circ}$) film is as narrow as the \ch{SrTiO3} substrate ($\Delta\omega\approx$ 0.007$^{\circ}$), suggesting high crystalline quality of the grown layers (Fig. \ref{fig:XRD}c). The films grown on \ch{GdScO3} and \ch{DyScO3} both show comparable FWHM values, along both orthogonal in-plane directions of the substrate (Fig. S5 and Fig. S6).

Thermodynamic analysis was applied to compare the out-of-plane lattice constant \textit{c}$_{KTO}$ of commensurately strained \ch{KTaO3} films on different mismatched substrates, using the paraelectric and ferroelectric ground states of \ch{KTaO3} (Fig. \ref{fig:XRD}d). The solid lines are calculated using a single set of coefficients (Table S2-3). The out-of-plane lattice constant of the strained \ch{KTaO3} films were determined from the 00$\mathcal{l}$ ($\mathcal{l}$=1,2,3) reflection peaks and the full diffraction spectra, described in the Supporting Information Section III. The calculated out-of-plane lattice constant, expected for a commensurately strained \ch{KTaO3} thin film on \ch{SrTiO3} at room temperature is 4.028 \AA, assuming a paraelectric ground state. XRD resolves an out-of-plane lattice constant of  4.063 $\pm$ 0.005 \AA~ for the \ch{KTaO3} film, grown commensurate to the \ch{SrTiO3} (001) substrate. The discrepancy between the measured and calculated out-of-plane lattice parameters could be due to a structural phase transition with epitaxial strain. Figure \ref{fig:XRD}d plots the calculated ferroelectric contribution to the out-of-plane lattice constant (orange region). Here, the measured out-of-plane lattice constant is consistent with the upper bound of the calculated ferroelectric contribution. The out-of-plane lattice constant becomes similar to the calculated value for the paraelectric ground state in partially relaxed films.\cite{schwaigert2023} We also note a previous effort to grow \ch{KTaO3} films commensurate to \ch{SrTiO3 } (001) using pulsed-laser epitaxy, reporting a 4.008 $\pm$ 0.0005 \AA~ out-of-plane lattice constant for \ch{KTaO3} films. \cite{tyunina2010} Our results differ significantly from the previous study. The improved composition and structural perfection, achieved through adsorption-controlled growth, suggest that our results reflect the intrinsic behavior of \ch{KTaO3}.

In contrast to the extended out-of-plane lattice spacing observed for the commensurately strained \ch{KTaO3} films grown on a \ch{SrTiO3} substrate, the 18 nm thick \ch{KTaO3} film on \ch{DyScO3} and the 18 nm thick film on \ch{GdScO3} appear consistent with the thermodynamic predictions for a paraelectric phase. Atomic force microscopy (AFM) micrographs (Fig. S4) reveal a root-mean-square (rms) roughness for the 9.3 nm thick \ch{KTaO3} on \ch{SrTiO3} of $\approx$ 0.5 nm, measured by taking a \SI{5}{\micro\square\metre} area as a reference. For the \ch{KTaO3} thin films grown on \ch{DyScO3} and \ch{GdScO3} the rms are 0.9 nm and 0.6 nm, respectively (see Fig. S5 and Fig. S6).

\begin{figure}[h]
\includegraphics[width=0.5\textwidth]{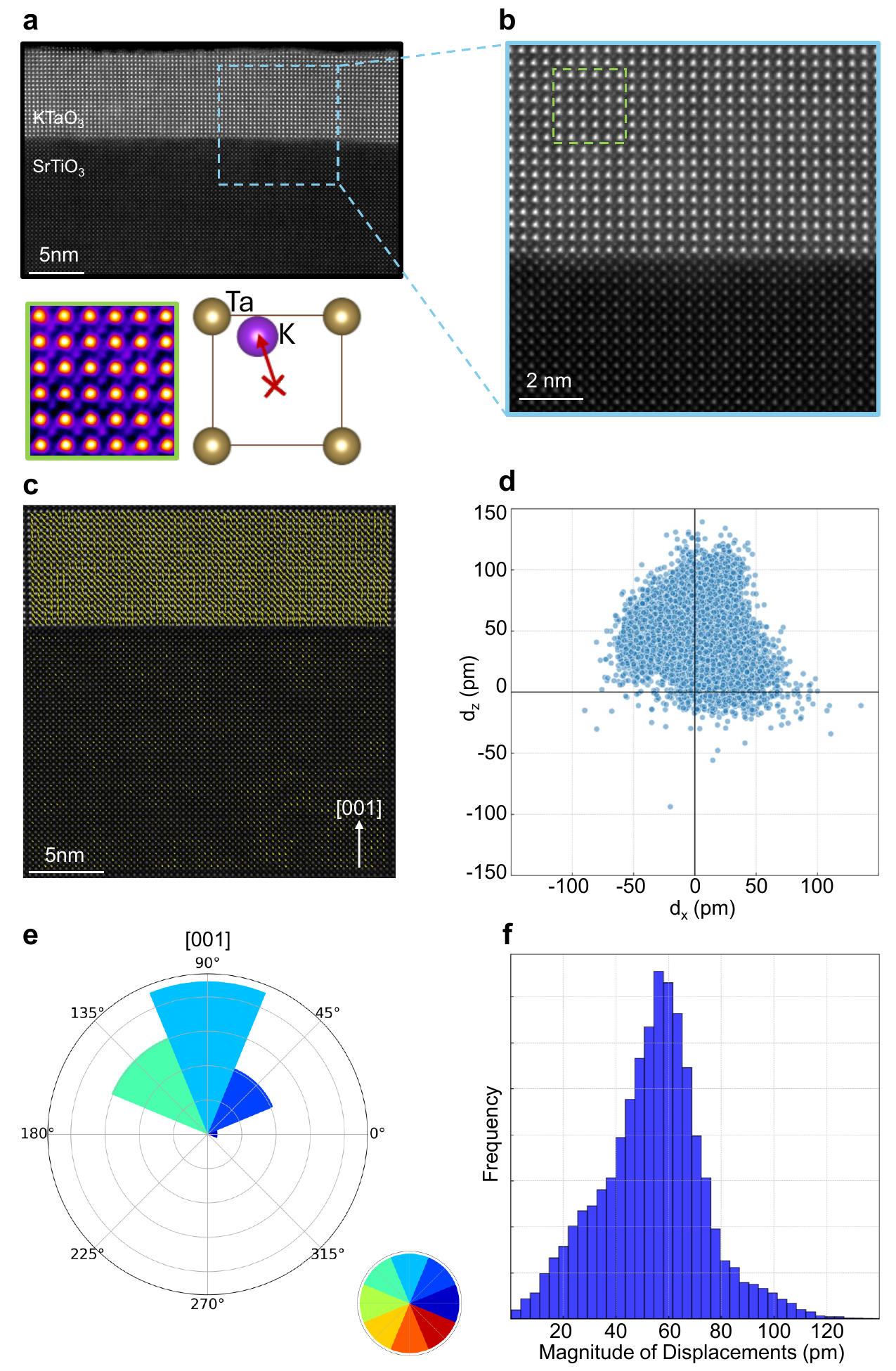}
\caption{\label{fig:STEM} \textbf{Probing the polar distortions at the atomic scale in epitaxially strained \ch{KTaO3}}. \textbf{a} HAADF-STEM image of the \ch{KTaO3} film grown on \ch{SrTiO3}(001).  \textbf{b} The interface structure is uniform throughout the sample, indicating a coherent strain between the film and the substrate. No extended defects are observed, which suggests high-quality epitaxial growth. \textbf{c} Displacement vector of the difference between the centeroid of four tantalum columns and potassium columns overlaid with HAADF-STEM image. \textbf{d} Tracing of the magnitude and direction of the displacements of atomic columns. \textbf{e} Polar histogram, based on more then 20,000 analyzed columns, shows the overall direction of displacements for the \ch{KTaO3} thin film grown on a \ch{SrTiO3} substrate. \textbf{f} Histogram of the magnitude of the polar displacements in potassium columns, based on the 20,000 columns analyzed.}
\end{figure}

HAADF-STEM was used to further investigate the polar distortions of the \ch{KTaO3} films. Figure \ref{fig:STEM}a and b display the HAADF-STEM image of the \ch{KTaO3} film grown on a \ch{SrTiO3} substrate. The interface between the \ch{KTaO3} film and the \ch{SrTiO3} substrate is sharp, and the film is strained to the substrate. This high-resolution image shows a coherent epitaxial interface, indicating a high-quality epitaxial relationship between the two materials. Figure \ref{fig:STEM}a also shows a zoomed-in region of the \ch{KTaO3} film, presented in false color, to highlight the displacement of the potassium column in relation to the center of the tantalum columns. The displacement is clearly visible, revealing the detailed atomic structure and the relative positions of the atomic columns. The accompanying schematic illustrates the assignment of the displacement vector. The polarization (displacement) vector is defined as the difference between the center of mass of four neighboring tantalum columns and the position of the potassium column obtained by Gaussian fitting. Displacements along the [001] direction (d$z$) and [100] direction (d$x$) are extracted, providing a detailed analysis of the atomic displacements in the film. Figure \ref{fig:STEM}c shows the displacement of the potassium column, indicated by the arrow. The polar displacements are negligible in the \ch{SrTiO3} substrate, but significant displacements are observed in the \ch{KTaO3} film. We attribute these displacements to strain-induced polarization in the \ch{KTaO3} films.

For a more robust analysis, we examined over 20,000 columns from multiple samples. Figure \ref{fig:STEM}d presents a scatter plot of the displacements along the [001] and the [100] crystallographic directions, revealing that the displacements are predominantly along the [001] direction. This indicates a preferential electric polarization along the out-of-plane direction, consistent with the theoretical prediction (Fig. \ref{fig:phasefield}). Figure \ref{fig:STEM}e provides a polar histogram of the vector angles, offering a visual representation of the displacement directions. The reference circles in the polar histogram indicate the frequency of displacements in each direction, with each segment showing the number of displacements occurring within a specific angular range. This visualization helps to confirm that the displacements are primarily aligned along the [001] direction. While the film on \ch{DyScO3} (Fig. S9) shows a slight net out-of-plane polar displacement, the film on \ch{GdScO3} (Fig. S10) does not have a net out-of-plane polar distortion (See Supporting Information V). Here, the displacements are only correlated on small length scale .

Figure \ref{fig:STEM}f exhibits the distribution of the potassium column polar displacement magnitude. The average magnitude of the polar displacements is 54.08 $\pm$ 19.75 pm. This distribution highlights the significant atomic displacements occurring in the \ch{KTaO3} film due to epitaxial strain. For small distortions, the change in polarization is linear to the displacements. We estimate the spontaneous polarization ($P_s$) using the following relationship:

\begin{equation}
\label{eq:BEC}
    P_S  = \frac{e}{V_{KTO}} \sum_{i} Z^*_i\times u_i,
\end{equation}
where e is the elementary charge, $V_{KTO}$ is the unit cell volume of \ch{KTaO3}, $Z^*_i$ the Born-effective charge (BEC) tensor coefficient and $u_i$ is the polar atomic distortions. The calculated BEC tensor can be found in Supporting Information Section VI. Equation \ref{eq:BEC} sums over the contributions of all the ions in the unit cell to the spontaneous polarization. Here, we only consider the polar distortions from potassium columns, resolved from HAADF-STEM, which estimates the room temperature spontaneous polarization of $P_s$ = 18 $\pm$ \SI{6}{\micro\coulomb\per\square\centi\metre}. 

\subsection*{Probing the emergence of Ferroelectricity}
\begin{figure}[h]
\includegraphics[width=\textwidth]{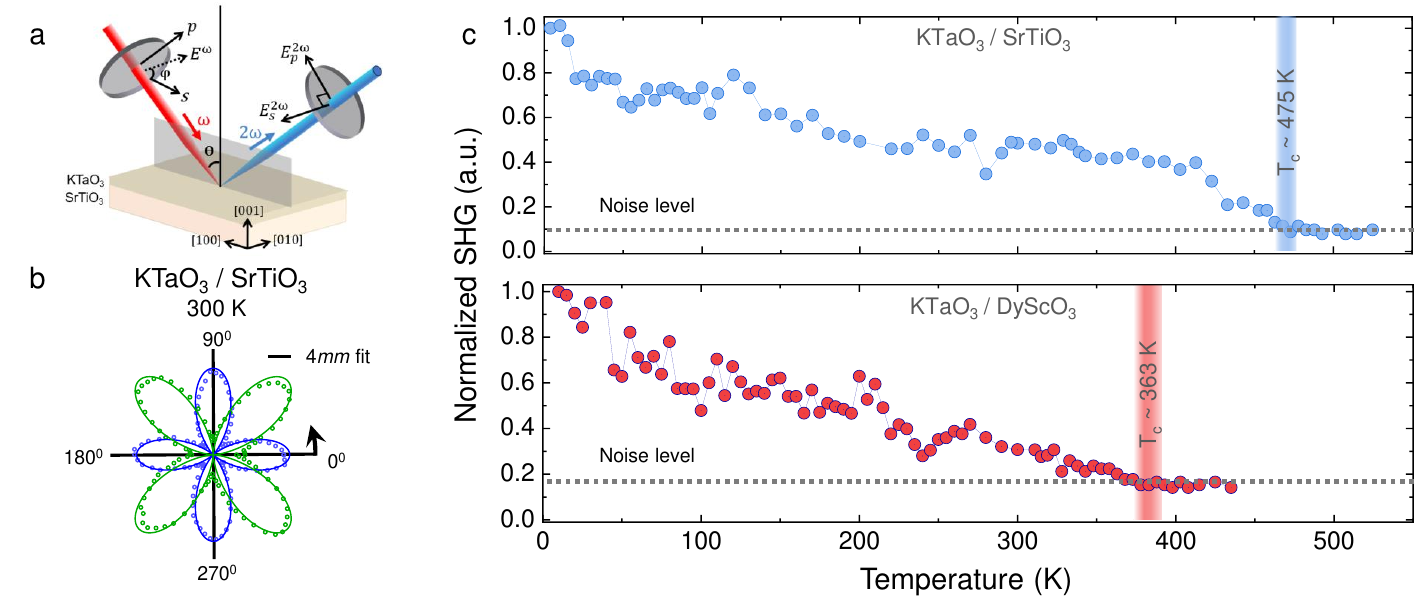}
\caption{\label{fig:SHG} \textbf{Emergence of the polar \textit{P}4\textit{mm} space group in epitaxially strained \ch{KTaO3}}. \textbf{a} Schematic of SHG setup in reflection geometry.\textbf{ b} Polar plots of SHG intensity (radius) versus fundamental
polarization (azimuth).\textbf{ c}  Normal and oblique incidence SHG intensity with temperature (10-550 K), shown for the 9.3 nm thick   \ch{KTaO3} film grwon on \ch{SrTiO3} and a 18 nm thick \ch{KTaO3} film grown on \ch{DyScO3}.}
\end{figure}

SHG measurements as a function of temperature were performed for the previously described \ch{KTaO3}/\ch{SrTiO3}, \ch{KTaO3}/\ch{DyScO3}, and \ch{KTaO3}/\ch{GdScO3} samples. A schematic of the SHG setup used for the experiment is illustrated in Fig. \ref{fig:SHG}a. SHG detects a signal that exceeds that of \ch{SrTiO3} substrate, indicating the long-range breaking of inversion symmetry. By rotating the polarization of the incident beam, polar plots can be generated, corresponding to \textit{p} and \textit{s}-polarized SHG light reflected off the sample. The derivation of the fitting model is further elaborated in the Supporting Information VII. For the largest in-plane compressive strain \ch{KTaO3}/\ch{SrTiO3} sample (-2.1\%) the phase transition temperature is at \SI{475}{K}, while for the \ch{KTaO3}/\ch{DyScO3} (-0.9\%) samples a polar phase is detected below 360 K. For temperatures below $T_C$, the strained \ch{KTaO3} films display the same tetragonal average symmetry as shown in Fig. \ref{fig:SHG}b, suggesting the absence of any other symmetry-lowering phase transitions.

If \ch{KTaO3} films on \ch{SrTiO3} exceed the critical thickness, resulting in strain relaxation, no SHG intensity beyond that of the substrate is detected (see Supporting Information VIII). Rotational anisotropy SHG results are successfully fitted to the 4$mm$ point group, further solidifying the emergence of a polar phase (Fig. \ref{fig:SHG}b), consistent with HAADF-STEM and XRD measurements. Although SHG confirms the emergence of spontaneous electric polarization, it is necessary to exhibit switching of polarization to confirm ferroelectricity. 

 We note the discrepancy between thermodynamic calculations and experimental evidence, XRD (Fig. 2d), HAADF-STEM (Fig. S8), and SHG (Fig. 4c) for the \ch{KTaO3} films grown on \ch{GdScO3}. Here, thermodynamic calculations suggest the emergence of a polar phase at around 100 K, while SHG does not show signal at any temperatures down to 4 K. To further support this observation, SHG intensity was measured as a function of incident beam fluence, which shows no measurable difference between \ch{KTaO3}/\ch{GdScO3} and a bare \ch{GdScO3} substrate at room and low temperatures (Fig. S11c). The out-of-plane lattice parameter from XRD (Fig. 2d) and  polarization mapping by HAADF-STEM (Fig. S10) also do not detect any sign of correlated polar distortions in \ch{KTaO3}/\ch{GdScO3} sample at room temperature. These results suggest that a threshold strain must be applied to stabilize the polar phase. This could also be observed in DFT calculations (Fig. 1b). The critical epitaxial strain of the ferroelectric transition calculated in Figure 1b depends on the exchange and correlational functional choices. The DFT results, however, suggests that a critical strain is needed for ferroelectric transition.

\begin{figure}[h]
\includegraphics[width=\textwidth]{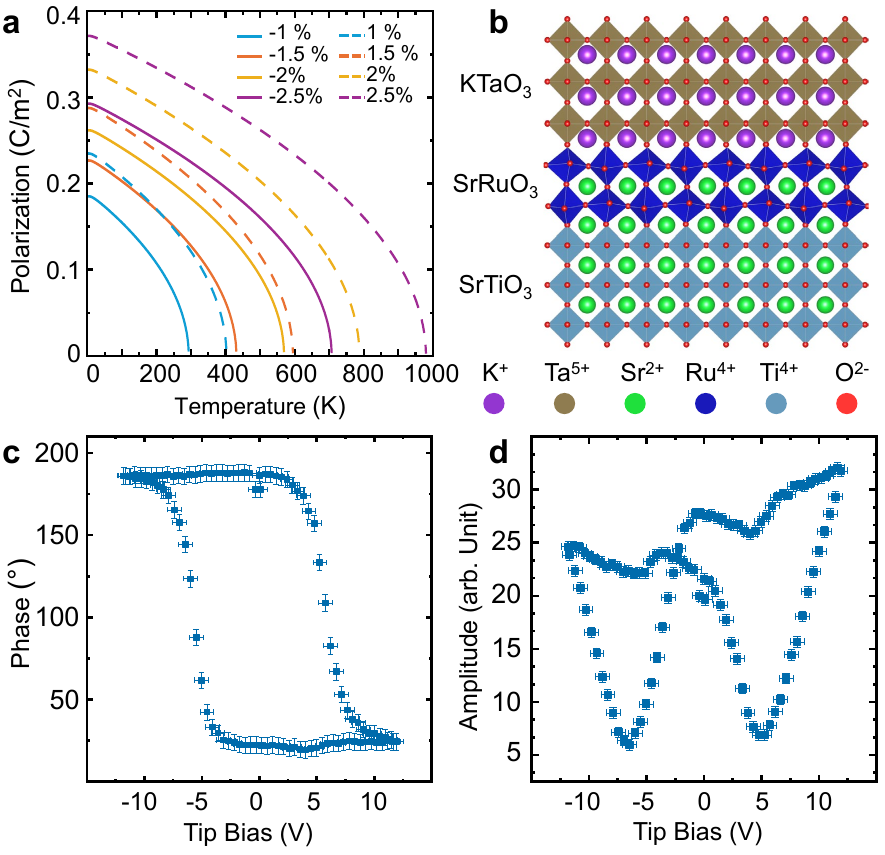}
\caption{\label{fig:Loop} \textbf{Switching of spontaneous electric polarization in epitaxially strained \ch{KTaO3}}. \textbf{a}  Thermodynamic calculation of the polarization as a function of epitaxial strain and temperature. \textbf{b} Schematic of the epitaxial heterostructure used for PFM measurements.\textbf{c} PFM switching using the \ch{KTaO3}/\ch{SrRuO3}/\ch{SrTiO3} sample showing a upward build-in field. \textbf{d} Magnitude of the piezo response.}
\end{figure}

Figure \ref{fig:Loop} shows phase-field simulation results, estimating the spontaneous polarization in \ch{KTaO3} with strain and temperature. Under epitaxial strain of –2.1\%, as imposed by coherent growth on \ch{SrTiO3}, the spontaneous polarization reaches $\sim$ \SI{26}{\micro\coulomb\per\square\centi\metre} at liquid helium temperature. The predicted room temperature electric polarization is $\sim$\SI{19}{\micro\coulomb\per\square\centi\metre} which is consistent with the STEM-resolved polarization (18 $\pm$ \SI{6}{\micro\coulomb\per\square\centi\metre}) using Eq. \ref{eq:BEC}, suggesting that the other contributions to electric polarization, e.g., anion distortions, are negligible.

To determine whether this strain-induced electric polarization is switchable under applied electric field, a capacitor stack was fabricated: \ch{KTaO3} grown on a conductive \ch{SrRuO3} bottom electrode on \ch{SrTiO3} (001). For structural details, see Supporting Information III. Local electromechanical measurements using piezoresponse force microscopy (PFM) were used to reveal ferroelectric switching and the hysteresis loop, instead of conventional PUND techniques, due to the leakage current that often dominates electrical measurements in ultra-thin layers.\cite{dawber2005physics} Figure \ref{fig:Loop} illustrates the heterostructure used for PFM: \ch{KTaO3}/\ch{SrRuO3}/\ch{SrTiO3}, where \ch{SrRuO3} serves as an epitaxial bottom electrode. Clear 180° phase reversal for out-of-plane PFM response at an arbitrary location on the capacitors of \ch{KTaO3}/\ch{SrRuO3}/\ch{SrTiO3} is shown in Fig.  \ref{fig:Loop}b. At a quasi-static frequency of 1.1 Hz standard ferroelectric $d_{33}$-V is observed with coercive voltages of $\pm$ 5.7 V. The error bars represent the average response over the 10 cycles of a ``pulse-mode" triangular wave.  The butterfly-shaped amplitude curve confirms ferroelectric switching (Fig. \ref{fig:Loop}c). Additionally, \ch{KTaO3}/\ch{SrRuO3}/\ch{DyScO3} capacitor heterostructure was investigated using PFM. A 180° phase reversal for the out-of-plane PFM response is observed (Fig. S12). This result is consistent with thermodynamic calculations (Fig. 1a) and SHG (Fig. 4b), exhibiting an above room temperature polar transition. The \ch{KTaO3} films grown on \ch{DyScO3}, however, do not demonstrate an extended XRD lattice constant (Fig. 2d) and HAADF-STEM shows a small net out-of-plane polar distortion (Fig. S9). Furthermore, the SHG transition appears smeared out, compared to the \ch{KTaO3} films grown on \ch{SrTiO3} (Fig. 4). This could be due to the presence of polar nano-regions and an order-disorder ferroelectric transition which has been reported in strained \ch{SrTiO3} films. \cite{xu2020strain, PhysRevLett.125.087601, salmani2020polar}. Finally, we do not observe any polar switching behavior in similar heterostructures grown on \ch{GdScO3}.

In conclusion, high-quality \ch{KTaO3} thin films were coherently strained to various substrates exerting a compressive in-plane strain ranging from -0.5\% on \ch{GdScO3} up to -2.1\% on \ch{SrTiO3}. The emergence of a tetragonal polar $P4mm$ phase is predicted by DFT and thermodynamic calculations and experimentally confirmed by diffraction, SHG, and HAADF-STEM. We demonstrated the switching of the emergent spontaneous electric polarization using PFM. The room temperature phase-field spontaneous polarization (\SI{19}{\micro\coulomb\per\square\centi\metre}) is consistent with the experimentally resolved value (18 $\pm$ 6 \unit{\micro\coulomb\per\square\centi\metre}) using HAADF-STEM atomic polar distortions and calculated BEC tensor. We have thus revealed a ferroelectric ground state in epitaxially strained \ch{KTaO3}. 

Given the recent discovery of unconventional superconductivity at $\text{KTaO}_3$ interfaces \cite{liu2021,kim2024electronic, arnault2023, al2023enhanced} and the ferroelectric enhancement of superconductivity in $\text{SrTiO}_3$,\cite{ahadi2019enhancing, PhysRevMaterials.3.091401} tunable ferroelectricity in $\text{KTaO}_3$ provides a platform for evaluating theoretical proposals that link superconductivity to a proximal ferroelectric state. \cite{yuan2022supercurrent, PhysRevB.97.144506, PhysRevB.105.224503}

\section{Methods}
\textbf{Thin film synthesis and characterization}
Epitaxial \ch{KTaO3} films were grown in a modified Vecco GEN 10 MBE system. The substrate was heated using a 10 $\mu$m \ch{CO2}-laser, Epiray GmbH (THERMALAS Substrate Heater). Films were grown by co-deposition of potassium, tantalum, and ozone at a substrate temperature of 650 °C, measured using an optical pyrometer operating at a \SI{7.5}{\micro\metre} wavelength.  Here, all the films were grown in an adsorption-controlled regime with a K:Ta flux ratio of approximately 10:1.  A mixture of  ozone and oxygen  (10 \% \ch{O3} + 90 \% \ch{O2}) was used as the oxidant. The films were grown at an oxidant background pressure of $1\times10^{-6}$ Torr. Typical fluxes for the sources were (4-7)$\times 10^{12}$ atoms/$\text{cm}^2$/s for the tantalum source and (4-7)$\times 10^{13}$ atoms/$\text{cm}^2$/s for the potassium source, determined by a quartz crystal microbalance (QCM), with an accuracy of about $\pm$ 15 \%. Co-deposition with these fluxes results in a \ch{KTaO3} film growth rate of about 0.03 \AA/s. Additional MBE growth details are described elsewhere.\cite{schwaigert2023} 

X-ray diffraction (XRD), X-ray reflectometry (XRR), and reciprocal space mapping (RSM) measurements were carried out using a PANalytical Empyrean diffractometer with Cu K$\alpha_{1}$ radiation. The raw XRR spectra were analyzed using the PANalytical X´Pert Reflectivity software package and the layer thickness was derived from a fast Fourier transform (FFT) after manually defining the critical angle to account for refractive effects. \textit{In situ} reflection high-energy electron diffraction (RHEED) patterns were recorded using KSA-400 software and a Staib electron source operated  at  14 kV and a filament current of 1.5 A. The morphology of the film surface was characterized using an Asylum Cypher ES environmental AFM. Piezoforce microscopy (PFM) was conducted on an Asylum Cypher AFM using Dual AC Resonance Tracking (DART) mode.

\textbf{Transmission electron microcopy}
Cross-sectional scanning transmission electron microscopy (STEM) samples were prepared using standard lift-out process using a Thermo Fisher Scientific Helios G4UX focused ion beam with the final milling voltage of 2 kV for the gallium ions. A Thermo Fisher Scientific Spectra 300 X-CFEG, operating at 200 kV with a convergence angle of 30 mrad and a high-angle annular dark-field (HAADF) detector with an angular range of 60-200 mrad, was used to collect atomic-resolution HAADF-STEM images. STEM energy-dispersive X-ray spectroscopy (EDX) data were collected using a steradian Dual-X EDX detector with a probe current of 100 pA. The noise of the STEM-EDX spectrum was reduced by the application of principal component analysis.

\textbf{Second-Harmonic Generation}
(SHG) measurements were made with a 800 nm fundamental wavelength pulsed laser line from a  Spectra-Physics Solstice Ace laser focused on the sample, while second-harmonic light generated at 400 nm was measured reflecting off the samples. The repetition rate of the laser was 1 kHz, while the beam size at focus was 16 $\mu$m  All measurements were done in reflection geometry. For low temperature measurements a Janis 30 gas flow cryostat was used, while for high temperature measurements a home made heater was employed.

\textbf{Density functional theory}
The atomic structure relaxation, phonon, and Berry phase calculations were performed using first-principles density functional theory within the generalized gradient approximation (GGA) as implemented in the PBEsol functional \cite{pbesol} by the Vienna Ab initio Simulation Package (VASP) \cite{vasp} with the ion-electron interaction described by the projector augmented wave method \cite{paw} including the spin-orbit coupling (SOC) effects. We employed an energy cutoff of 700 eV, Gaussian smearing of 0.005 eV, and electronic energy tolerance of $10^{-8}$ eV for the total energy convergence. The ionic relaxations were performed with a force tolerance of $10^{-3}$ Å/eV and an electronic momentum \textit{k}-point mesh of $24 \times 24 \times\ 24$. Phonon and Berry phase calculations were performed with an electronic momentum \textit{k}-point mesh of $16 \times 16 \times 16$. Phonon calculations were analyzed using the Phonopy code \cite{phonopy-phono3py-JPCM}.

\textbf{Thermodynamics}
Using Landau-Ginzburg-Devonshire theory for ferroelectrics, the elastic Gibbs free energy density \textit{g}, written as a power series of the polarization near the phase transition (Supporting Information discussion II), describes the ferroelectric phase transition.\cite{haun1987} The ferroelectic transition temperature can be calculated by solving equations 1 and 2, which allows the construct of the phase transition boundaries between the paraelectric and ferroelectric phases in the temperature-strain phase diagram. Since there is some variation in the data collected on the electrostrictive coefficients and elastic constants from the literature, we list the values by different groups in Supporting Information Tables S2 and S3 and include the regions indicating the range of ferroelectric transition to account for the spread in these values exhibited in Fig. 1a.

 \begin{acknowledgments}
K.A is supported by the U.S. National Science Foundation under Grant No. NSF DMR-2408890. K.A also acknowledges a seed grant from the Center for Emergent Materials, an NSF MRSEC, under Grant No. DMR-2011876. This material is based upon work supported by the National Science Foundation (Platform for the Accelerated Realization, Analysis, and Discovery of Interface Materials (PARADIM)) under Cooperative Agreement No. DMR-2039380. S.H., A.R., L-Q.C, and V.G. acknowledge support from the National Science Foundation grant  DMREF DMR-2011839 for SHG characterization and phase-field modeling. A.R. acknowledges the support of the National Science Foundation Graduate Research Fellowship Program under Grant No. DGE1255832. This work made use of a Helios FIB supported by NSF (Grant No. DMR-1539918) and the Cornell Center for Materials Research (CCMR) Shared Facilities, which are supported through the NSF MRSEC Program (Grant No. DMR-1719875). The authors acknowledge Steve Button for substrate preparation. We gratefully acknowledge Dasol Yoon and Xiyue Zhang for providing EDX analysis code.
\end{acknowledgments}
\appendix
\section*{AUTHOR DECLARATIONS}
\textbf{Conflict of Interest}
The authors have no conflicts to disclose.
\section*{Data Availability Statement}
The data that support the findings of this study are available within the article. Additional data related to film growth and structural characterization by XRD and STEM \textbf{are available at https://doi.org/10.34863/xxxx} and data related to DFT calculations are available at https://doi.org/xxxx.

\bibliography{aipsamp.bib}


\section*{Supplementary material}

Supporting Information: Above Room Temperature Ferroelectricity in Epitaxially Strained $KTaO_3$

TABLE S3. Elastic compliances in units of $10^{-12}$ $Pa^{-1}$

|  | Ref 9 | Ref 10 | Ref 10 | Ref 11 |
|---|---|---|---|---|
| $s_{11}$ | 2.555 | 2.47 | 2.17 | 2.36 |
| $s_{12}$ | -0.493 | -0.325 | -0.254 | -0.307 |
| $s_{44}$ | 9.174 | 5.9 | 5.076 | 11.69 |

## III. ADDITIONAL SAMPLE CHARACTERIZATION OF $KTaO_3$ FILMS

High-resolution X-ray diffraction (XRD; Panalytical Emperyean) using Cu $K_{\alpha 1}$ radiation was performed on $KTaO_3$ thin films grown on different substrates. Figure S3 show $\theta$-$2\theta$ scans of phase-pure epitaxial (001)-oriented $KTaO_3$ films on three different substrates ($SrTiO_3$, $DyScO_3$, and $GdScO_3$). In total 9.3 nm of $KTaO_3$ were deposited on $SrTiO_3$ (001) and 18 nm on each of $DyScO_3$ $(110)_o$ and $GdScO_3$ $(110)_o$. The presence of Laue fringes indicate that each film has a well-defined thickness i.e., smooth interfaces between film and substrate and between film and air. These Laue fringes around the thin films peak were used to calculate the thickness and were equal to the thickness extracted from HAADF-STEM. The in-plane lattices constants of the strained $KTaO_3$ films shown in Fig. 2d were determined from the $00\ell$ ($\ell$=1,2,3) reflection peaks. Supplementary Figure S5a shows the reciprocal space

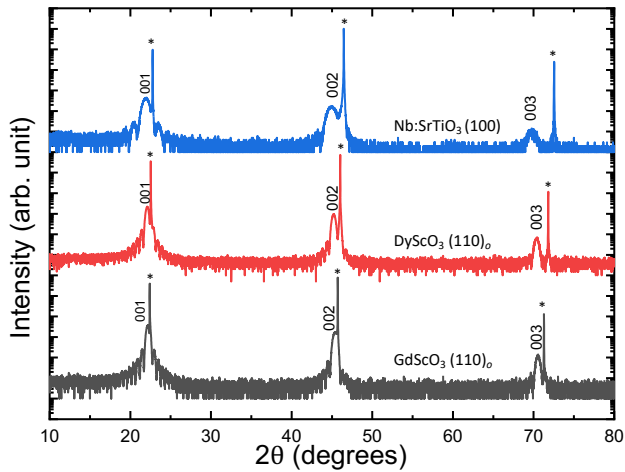

Fig. S3. $\theta$-$2\theta$ XRD scan of $KTaO_3$ films on (001) $SrTiO_3$ (blue), $(110)_o$ $DyScO_3$, (red) and $(110)_o$ $GdScO_3$ (grey) substrates.

maps (RSM) around the 103 $KTaO_3$ and 332 $DyScO_3$ $(110)_o$. The film is commensurately

Supporting Information: Above Room Temperature Ferroelectricity in Epitaxially Strained $KTaO_3$

strained to the substrate. The FWHM of the rocking curves of film (0.009°/0.003°) and substrate (0.008°/0.003°) are comparable, suggesting high crystalline quality (Supplementary Fig. S5b,d). The rocking curves were measured along the [001] and [1$\bar{1}$0] in-plane directions of the $DyScO_3$ substrate, respectively. This is the narrowest FWHM ever reported for a $KTaO_3$ film. Figure S5c,f shows AFM images at different magnifications. The root-mean-square (rms) roughness is 0.8 nm taking a 5 µm² area as the reference.

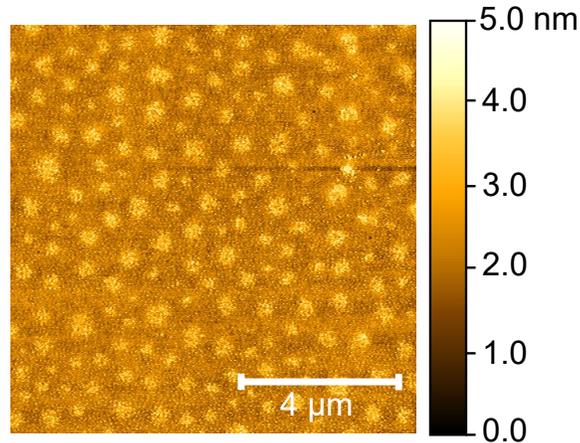

Fig. S4. AFM images of the untreated 9.3 nm thick $KTaO_3$ film shown through Figs. 2-4. This film exhibits an rms roughness of 0.8 nm over the whole 5 µm² scan.

Figure S6**a** shows the RSM around the 103 $KTaO_3$ and 332 $GdScO_3$ reflection and suggests that the film is commensurately strained to the substrate. The FWHM of the rocking curves of film (0.009°/0.008°) and substrate (0.017°/0.016°) comparable suggesting high crystalline quality (Fig. S5b,d). Figure S6 shows the AFM images at different magnifications with a rms of 0.6 nm taking a 5 µm² area as the reference.

Supporting Information: Above Room Temperature Ferroelectricity in Epitaxially Strained KTaO$_3$

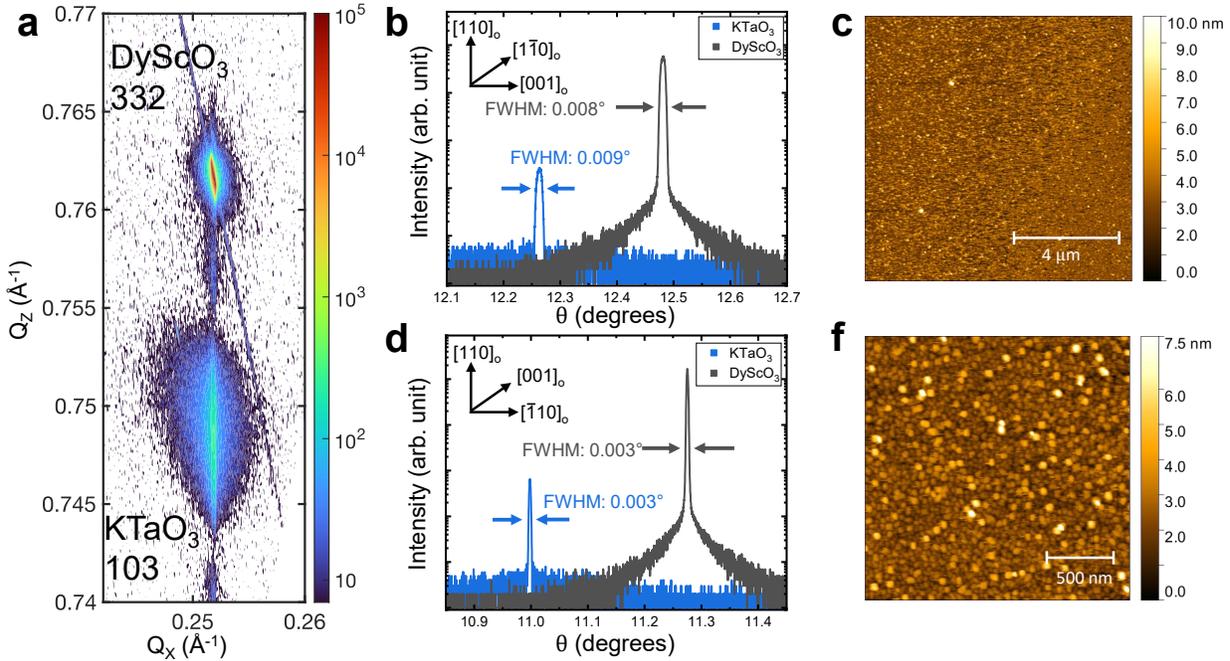

Fig. S5. **18 nm KTaO$_3$ on DyScO$_3$.** **a** RSM of the KTaO$_3$ 103 and DyScO$_3$ 332 peaks. **b and d** Overlaid rocking curves of the KTaO$_3$ 001 and GdScO$_3$ 110 peaks, showing comparable FWHMs, indicating low out-of-plane mosaicity ($\Delta\omega \approx 0.009°$ and $0.003°$ along the two orthogonal in-plane directions of the substrate.) **c,f** AFM images at different magnifications. The root mean square (rms) roughness over the $5\,\mu m^2$ scan is 0.9 nm.

9

Supporting Information: Above Room Temperature Ferroelectricity in Epitaxially Strained KTaO$_3$

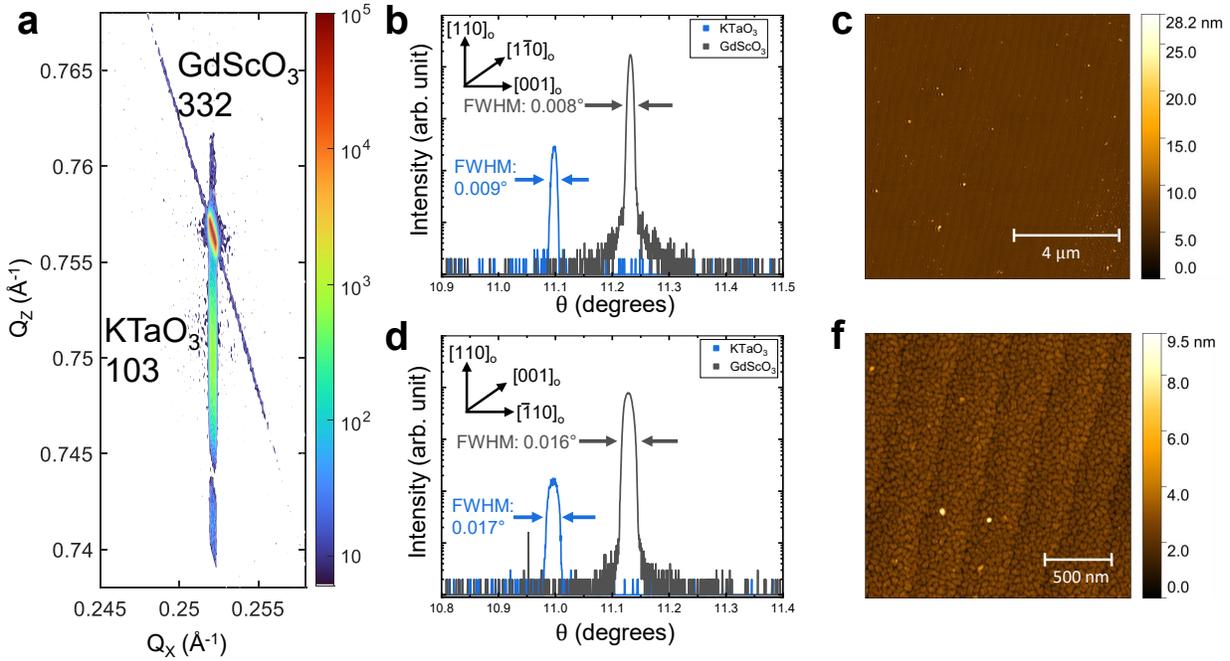

Fig. S6. **18 nm KTaO$_3$ on GdScO$_3$ a** RSM of the KTaO$_3$ 103 and GdScO$_3$ 332 peaks. **b and d** Overlaid rocking curves of the KTaO$_3$ 001 and GdScO$_3$ 110 peaks, showing comparable FWHMs, indicating low out-of-plance mosaicity ($\Delta\omega \approx 0.009°$ and $0.017°$ along the two orthogonal in-plane directions of the substrate.) **c,f** AFM images at different magnifications. The root mean square (rms) roughness over the $5\,\mu m^2$ scan is 0.6 nm.

Supporting Information: Above Room Temperature Ferroelectricity in Epitaxially Strained $KTaO_3$

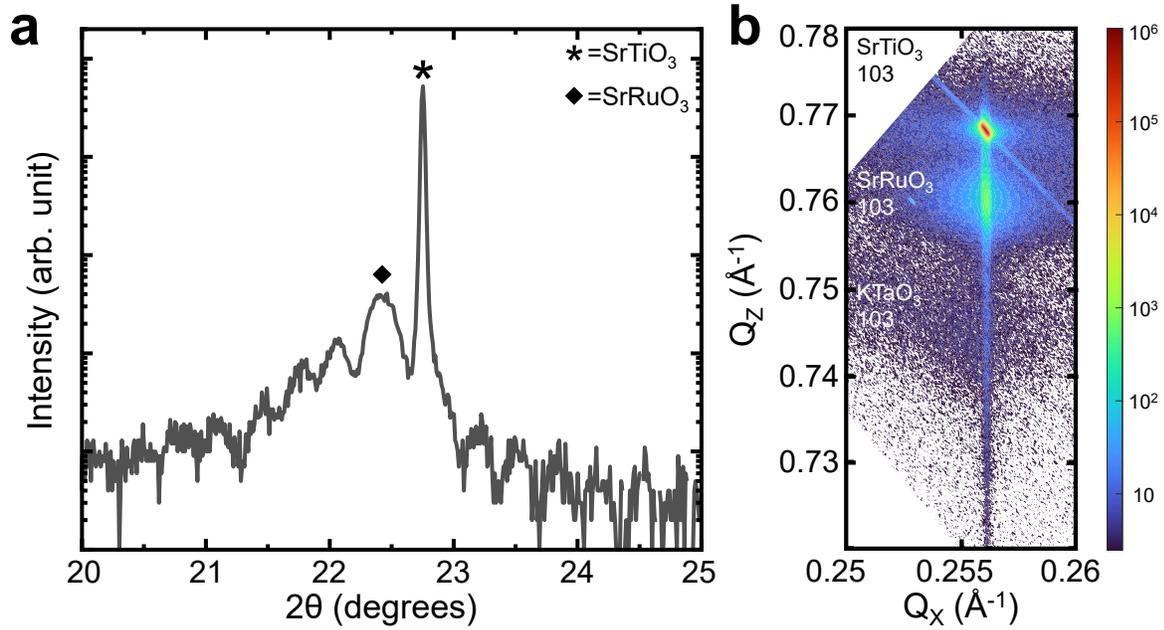

Fig. S7. **a** $\theta$-$2\theta$ XRD scan of $KTaO_3$ films on a 35 nm thick bottom electrode $SrRuO_3$ deposited on a $SrTiO_3$ substrate. **b** RSM of the $KTaO_3$ 103, $SrRuO_3$ 332, and $SrTiO_3$ 103 peaks.

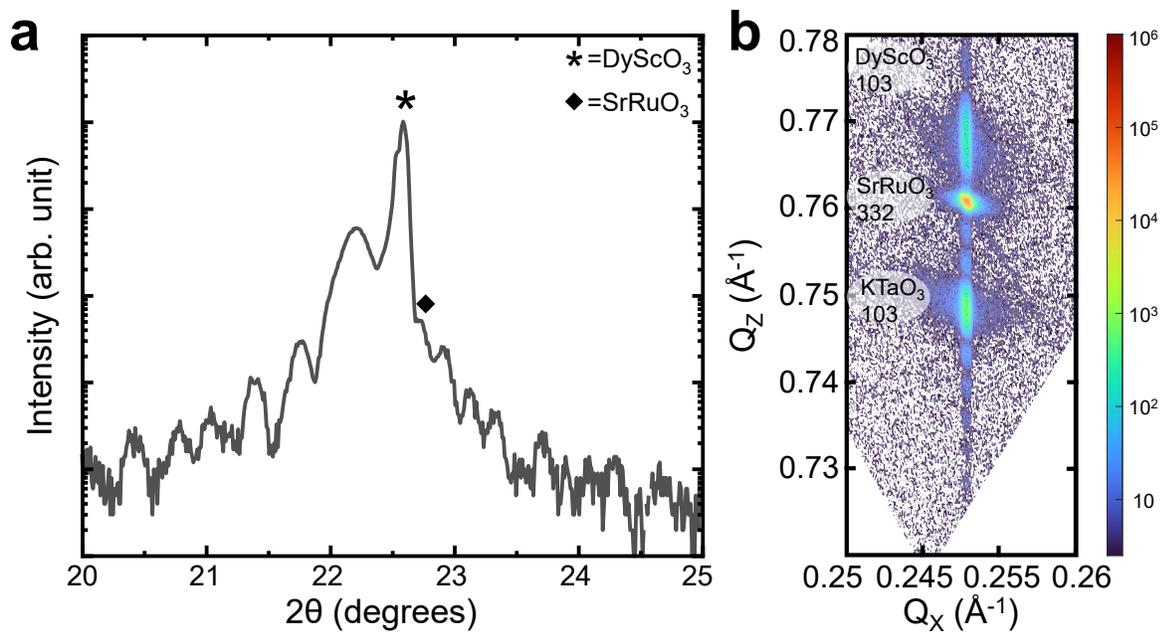

Fig. S8. **a** $\theta$-$2\theta$ XRD scan of $KTaO_3$ films on a 20 nm thick bottom electrode $SrRuO_3$ deposited on a $DyScO_3$ substrate. **b** RSM of the $KTaO_3$ 103, $SrRuO_3$ 332, and $DyScO_3$ 332 peaks.

11

Supporting Information: Above Room Temperature Ferroelectricity in Epitaxially Strained $KTaO_3$

## IV. THERMODYNAMIC ANALYSIS OF EFFECT OF STRAIN ON PARA AND FERROELECTRIC GROUND STATE OF $KTAO_3$

For a ferroelectric thin film epitaxially grown on a substrate, we may assume that the in-plane strains are fixed by the lattice parameter of the substrate and the out-of-plane components of stress are free to relax. Therefore, the in-plane components of strain ( $\varepsilon_1, \varepsilon_2$, and $\varepsilon_6$ ) and the out-of-plane components of stress ($\sigma_3, \sigma_4$, and $\sigma_5$ ) will be independent variables leaving, the in-plane components of stress ($\sigma_1, \sigma_2$, and $\sigma_6$ ) and the out-of-plane components of strain ( $\varepsilon_3, \varepsilon_4$, and $\varepsilon_5$ ) as dependent variables. While equation S1 describes the thermodynamic behavior using stress as the independent variable and strain as the dependent variable, we can transform this equation to be more suitable for ferroelectric thin films,

$$f(T, P, \varepsilon_1, \varepsilon_2, \sigma_3) = u - Ts - \sigma_3 \varepsilon_3 = g + \sigma_1 \varepsilon_1 + \sigma_2 \varepsilon_2 + \sigma_6 \varepsilon_6 \qquad (11)$$

To perform the Legendre transform, we must find $\sigma_1(T, P, \varepsilon_1, \varepsilon_2, \sigma_3)$, $\sigma_2(T, P, \varepsilon_1, \varepsilon_2, \sigma_3)$ and $\sigma_6(T, P, \varepsilon_1, \varepsilon_2, \sigma_3)$ . For simplicity we shall only consider the impact of normal strains. To do this we shall first calculate

$$\varepsilon_1(T, P, \sigma_1, \sigma_2, \sigma_3) = -\left(\frac{\partial g}{\partial \sigma_1}\right)_{T,P} = Q_{11}P_1^2 + Q_{12}(P_2^2 + P_3^2) + S_{11}\sigma_1 + S_{12}(\sigma_2 + \sigma_3) \qquad (12)$$

$$\varepsilon_2(T, P, \sigma_1, \sigma_2, \sigma_3) = -\left(\frac{\partial g}{\partial \sigma_2}\right)_{T,P} = Q_{11}P_2^2 + Q_{12}(P_1^2 + P_3^2) + S_{11}\sigma_2 + S_{12}(\sigma_1 + \sigma_3) \qquad (13)$$

Then algebraically solve for $\sigma_1(T, P, \varepsilon_1, \varepsilon_2, \sigma_3)$ and $\sigma_2(T, P, \varepsilon_1, \varepsilon_2, \sigma_3)$ which yields

$$\sigma_1 = \frac{\varepsilon_1 - (Q_{11}P_1^2 + Q_{12}(P_1^2 + P_3^2)) - (S_{12} - \frac{S_{12}^2}{S_{11}})\sigma_3 + \frac{S_{12}}{S_{11}}(Q_{11}P_2^2 + Q_{12}(P_1^2 + P_3^2) - \varepsilon_2)}{S_{11} - \frac{S_{12}^2}{S_{11}}}, \qquad (14)$$

$$\sigma_2 = \frac{\varepsilon_2 - (Q_{11}P_2^2 + Q_{12}(P_2^2 + P_3^2)) - (S_{12} - \frac{S_{12}^2}{S_{11}})\sigma_3 + \frac{S_{12}}{S_{11}}(Q_{11}P_1^2 + Q_{12}(P_2^2 + P_3^2) - \varepsilon_1)}{S_{11} - \frac{S_{12}^2}{S_{11}}}. \qquad (15)$$

Substituting in these expressions for $\sigma_1(T, P, \varepsilon_1, \varepsilon_2, \sigma_3)$ and $\sigma_2(T, P, \varepsilon_1, \varepsilon_2, \sigma_3)$ in equation 2.1 we obtain an expression that uses the in-plane strain and the out-of-plane stress as the

Supporting Information: Above Room Temperature Ferroelectricity in Epitaxially Strained KTaO$_3$

independent variables.

$$f(T, P, \varepsilon_1, \varepsilon_2, \sigma_3) = g(T, P, \sigma_1(T, P, \varepsilon_1, \varepsilon_2, \sigma_3), \sigma_1(T, P, \varepsilon_1, \varepsilon_2, \sigma_3), \sigma_3) + \sigma_1(T, P, \varepsilon_1, \varepsilon_2, \sigma_3)\varepsilon_1 + \sigma_2(T, P, \varepsilon_1, \varepsilon_2, \sigma_3)\varepsilon_2 \tag{16}$$

Where the out-of plane strain is found by taking the derivative of the free-energy with respect to the out-of plane stress

$$\varepsilon_3 = \left(\frac{\partial f}{\partial \sigma_3}\right)_{T,P,\varepsilon_1,\varepsilon_2} \tag{17}$$

Assuming we have a tetragonal C domain $P_i = (0, 0, P_3)$, we can find that the out-of plane strain is

$$\varepsilon_3 = \frac{(\varepsilon_1 + \varepsilon_2)S_{12}}{S_{11} + S_{12}} + \left(Q_{11} - \frac{2Q_{12}S_{12}}{S_{11} + S_{12}}\right)P_3^2 \tag{18}$$

Then, to obtain the lattice parameter one can use the relation

$$\varepsilon_3 = \frac{c - a^{eq}}{a^{eq}}, \tag{19}$$

where $a^{eq}$ is the equivalent cubic lattice parameter of a stress-free, polarization-free, cubic KTaO$_3$. Combining equation 20 and 21 we find,

$$c = a^{eq}[1 + \frac{(\varepsilon_1 + \varepsilon_2)S_{12}}{S_{11} + S_{12}} + (Q_{11} - \frac{2Q_{12}S_{12}}{S_{11} + S_{12}})P_3^2], \tag{20}$$

In the paraelectric phase, $P_i = (0, 0, 0)$, which yields

$$c = a^{eq}[1 + \frac{(\varepsilon_1 + \varepsilon_2)S_{12}}{S_{11} + S_{12}}], \tag{21}$$

and reduces to the solution derived from elasticity theory.

## V. STATISTICAL ANALYSIS OF ATOMIC DISPLACEMENTS

HAADF-STEM images of KTaO$_3$ thin film grown on DyScO$_3$ (Fig. S9) and on GdScO$_3$ (Fig. S10) substrates. Tracing the net magnitude of displacement of atomic columns of the HAADF-STEM image. While the film on DyScO$_3$ shows a slight net out-of-plane polar displacement, the film on GdScO$_3$ does not have net out-of-plane polar distortion. Here, the displacements are only correlated on a small length scale.

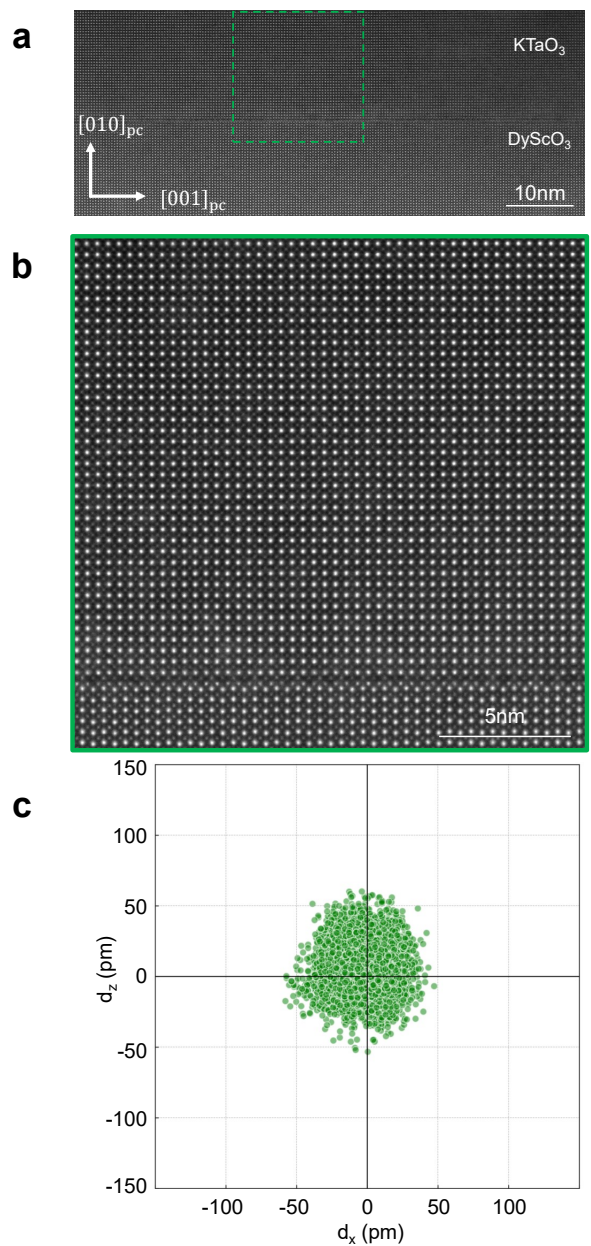

Fig. S9. **a,b** HAADF-STEM images of KTaO$_3$ thin film grown on DyScO$_3$ substrate. **c** Tracing the magnitude of displacement of atomic columns of the HAADF-STEM image.

Supporting Information: Above Room Temperature Ferroelectricity in Epitaxially Strained $KTaO_3$

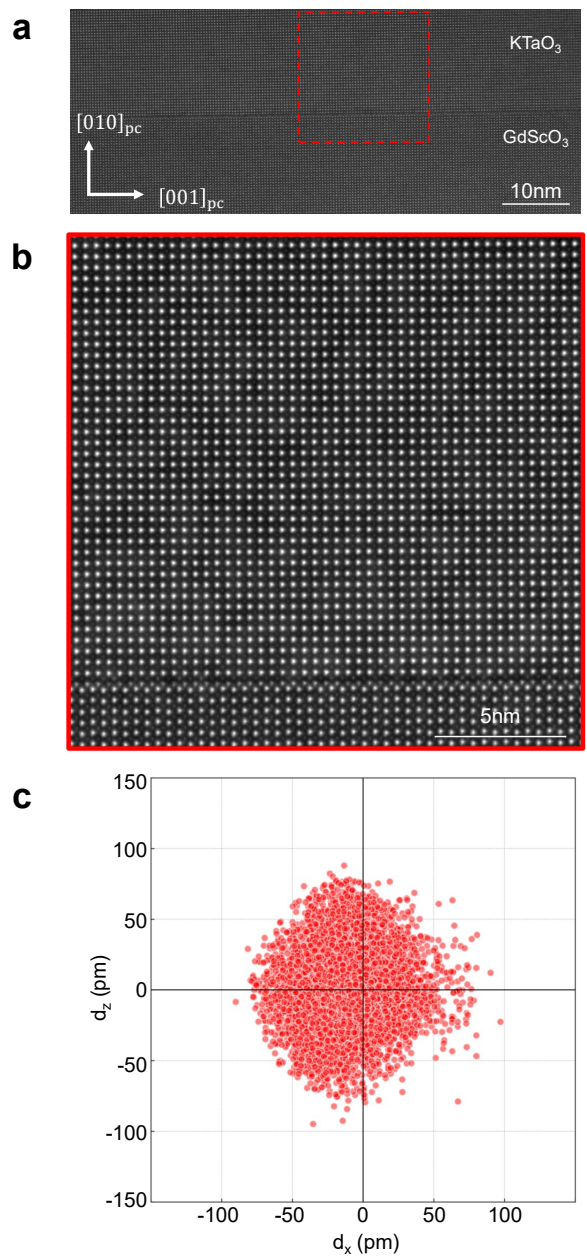

Fig. S10. **a,b** HAADF-STEM images of $KTaO_3$ thin film grown on $GdScO_3$ substrate. **c**vTracing the magnitude of displacement of atomic columns of the HAADF-STEM image.

15

Supporting Information: Above Room Temperature Ferroelectricity in Epitaxially Strained $KTaO_3$

# VI. BORN EFFECTIVE CHARGES FROM FIRST PRINCIPLES

Born effective charges were calculated from the linear response within the VASP package. The Born effective charge $Z^*$ matrices for each strain are shown in Table S4.

TABLE S4. Born effective charge $Z^*$ matrices in $|e|$ for each element with different strain.

| element | $KTaO_3$ (bulk) | $GdScO_3$ (-0.5 %) | $DyScO_3$ (-0.9 %) | $SrTiO_3$ (-2.1 %) |
|---|---|---|---|---|
| K | $\begin{pmatrix} 1.21 & 0.00 & 0.00 \\ 0.00 & 1.21 & -0.03 \\ 0.00 & 0.00 & 1.21 \end{pmatrix}$ | $\begin{pmatrix} 1.27 & 0.00 & 0.00 \\ 0.00 & 1.19 & 0.00 \\ 0.00 & 0.04 & 1.19 \end{pmatrix}$ | $\begin{pmatrix} 1.23 & 0.00 & 0.00 \\ 0.00 & 1.19 & 0.02 \\ 0.00 & 0.03 & 1.20 \end{pmatrix}$ | $\begin{pmatrix} 1.25 & 0.00 & 0.00 \\ 0.00 & 1.22 & 0.01 \\ 0.00 & 0.01 & 1.21 \end{pmatrix}$ |
| Ta | $\begin{pmatrix} 5.06 & 0.00 & 0.00 \\ 0.00 & 4.32 & 0.23 \\ 0.00 & -0.23 & 4.32 \end{pmatrix}$ | $\begin{pmatrix} 4.99 & 0.00 & 0.00 \\ 0.00 & 8.22 & 0.04 \\ 0.00 & -0.13 & 8.03 \end{pmatrix}$ | $\begin{pmatrix} 4.85 & 0.00 & 0.00 \\ 0.00 & 7.99 & 0.03 \\ 0.00 & -0.14 & 7.44 \end{pmatrix}$ | $\begin{pmatrix} 4.65 & 0.00 & 0.00 \\ 0.00 & 7.66 & 0.11 \\ 0.00 & -0.06 & 6.59 \end{pmatrix}$ |
| O | $\begin{pmatrix} -1.25 & 0.00 & 0.00 \\ 0.00 & -3.51 & -0.07 \\ 0.00 & -0.16 & -0.99 \end{pmatrix}$ | $\begin{pmatrix} -1.21 & 0.00 & 0.00 \\ 0.00 & -6.40 & -0.11 \\ 0.00 & -0.31 & -1.52 \end{pmatrix}$ | $\begin{pmatrix} -1.09 & 0.00 & 0.00 \\ 0.00 & -6.10 & -0.05 \\ 0.00 & -0.15 & -1.50 \end{pmatrix}$ | $\begin{pmatrix} -1.06 & 0.00 & 0.00 \\ 0.00 & -5.93 & -0.03 \\ 0.00 & -0.06 & -1.40 \end{pmatrix}$ |
| O | $\begin{pmatrix} -3.77 & 0.00 & 0.00 \\ 0.00 & -1.03 & -0.30 \\ 0.00 & 0.30 & -1.03 \end{pmatrix}$ | $\begin{pmatrix} -3.83 & 0.00 & 0.00 \\ 0.00 & -1.46 & -0.01 \\ 0.00 & 0.15 & -1.59 \end{pmatrix}$ | $\begin{pmatrix} -3.82 & 0.00 & 0.00 \\ 0.00 & -1.60 & -0.02 \\ 0.00 & 0.09 & -1.52 \end{pmatrix}$ | $\begin{pmatrix} -3.70 & 0.00 & 0.00 \\ 0.00 & -1.53 & -0.01 \\ 0.00 & 0.07 & -1.41 \end{pmatrix}$ |
| O | $\begin{pmatrix} -1.25 & 0.00 & 0.00 \\ 0.00 & -0.99 & 0.16 \\ 0.00 & 0.07 & -3.51 \end{pmatrix}$ | $\begin{pmatrix} -1.22 & 0.00 & 0.00 \\ 0.00 & -1.55 & 0.16 \\ 0.00 & 0.25 & -6.11 \end{pmatrix}$ | $\begin{pmatrix} -1.18 & 0.00 & 0.00 \\ 0.00 & -1.48 & 0.02 \\ 0.00 & 0.18 & -5.62 \end{pmatrix}$ | $\begin{pmatrix} -1.14 & 0.00 & 0.00 \\ 0.00 & -1.42 & -0.08 \\ 0.00 & 0.03 & -4.98 \end{pmatrix}$ |

# VII. DERIVATION OF OPTICAL SECOND HARMONIC GENERATION (SHG) POLARIMETRY EXPRESSION FOR THE TETRAGONAL PHASE *4mm*

Symmetry analysis of the $KTaO_3$ thin films was performed through fitting SHG polarimetry to analytical model as described below. The orientation of the substrate with respect to lab coordinates $(x_1, x_2, x_3)$ are described in Fig. 4a. All measurements are done in reflection geometry.

Supporting Information: Above Room Temperature Ferroelectricity in Epitaxially Strained $KTaO_3$

The induced nonlinear polarisation, $P^{2\omega}$, and the incident electric field, $E^{\omega}$, is related in a second harmonic generation process as,

$$P_i^{2\omega} \propto d_{ijk} E_j^{\omega} E_k^{\omega} \tag{22}$$

where, the proportionality constants depend on the incident beam fluence, beam shape, and Fresnel coefficients at the film-air, film-substrate interfaces and thickness of the films []. $d_{ijk}$ is the nonlinear susceptibility tensor and depends on the symmetry of the material in the crystal physics. For tetragonal $4mm$ point group symmetry, $d_{ijk}$ has the following form in Voigt notation:

$$d_{ij} = \begin{pmatrix} 0 & 0 & 0 & 0 & d_{15} & 0 \\ 0 & 0 & 0 & d_{15} & 0 & 0 \\ d_{31} & d_{31} & d_{33} & 0 & 0 & 0 \end{pmatrix} \tag{23}$$

The linearly polarized incident electric field $E^{\omega}$ is rotated by $\varphi$ and hence has the form $E_{\omega} = (E_{\omega} \cos(\psi), E_{\omega} \sin(\psi), 0)$. Taking into account the oblique incidence measurement geometry, the electric field in the crystal axes coordinates can be written as $(E_0 \cos(\varphi) \cos(\theta), E_0 \sin(\varphi), -E_0 \cos(\varphi) \sin(\theta))$, where $\theta$ is the angle of incidence.

The induced polarization in the crystal physics coordinates, $i = (1, 2, 3)$, can be calculated through equation 3, which can be rotated back to $(x_1, x_2, x_3)$ coordinates to give the $p$ and $s$-polarized components of SHG ($p||X$ and $s||Y$). The $p$ and $s$-polarized SHG intensities can be expressed as follows:

$$\begin{aligned} I_p^{2\omega} &\propto \left(P_p^{2\omega}\right)^2 \propto \left((2d_{15} - d_{31} - d_{33})\cos[\varphi]^2 - 2d_{31}\sin[\varphi]^2\right)^2 \\ I_s^{2\omega} &\propto \left(P_s^{2\omega}\right)^2 \propto d_{15}^2 \sin[2\varphi]^2 \end{aligned} \tag{24}$$

## VIII. INCIDENT POWER DEPENDENCE SHG MEASUREMENTS

Figure S11 shows the SHG intensity as a function of incident laser power measured for strained and relaxed $KTaO_3$ films on $SrTiO_3$, $DyScO_3$, and $GdScO_3$ substrates. While strained films on $SrTiO_3$ and $DyScO_3$ substrates show SHG beyond bare substrate, the relaxed films and the films grown on $GdScO_3$ do not show SHG sginal beyond their respective bare substrates.

17

Supporting Information: Above Room Temperature Ferroelectricity in Epitaxially Strained KTaO$_3$

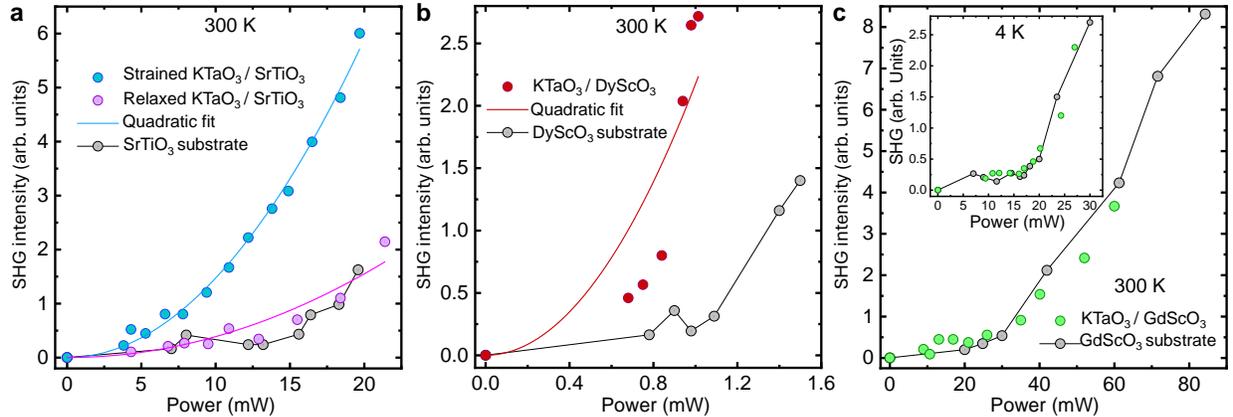

Fig. S11. **a**. SHG intensity as a function of incident laser power measured for strained and relaxed KTaO$_3$ films on SrTiO$_3$ substrate compared to similar measurement on a bare SrTiO$_3$ substrate. Thinner strained KTaO$_3$ film (9 nm thick) shows a prominent SHG response quadratically scaling with the incident fundamental laser power. Thicker relaxes KTaO$_3$ film (19 nm thick), however, shows no SHG response relative to the bare SrTiO$_3$ substrate. **b**. Quadratically scaling power depedent SHG response of strained KTaO$_3$ films grown on DyScO$_3$ substrate. **c**. Power dependent SHG measured on strained KTaO$_3$ film grown on GdScO$_3$ shows no SHG response different from measurements on bare GdScO$_3$ substrate.

## IX. PIEZO FORCE MICROSCOPY OF KTaO$_3$ ON DyScO$_3$

Figure S12 shows the KTaO$_3$/SrRuO$_3$/DyScO$_3$ capacitor hetero-structure is investigated using PFM. A 180° phase reversal for the out-of-plane PFM response is observed. The structural characterization details of this sample are discussed in Supplemental Discussion III.

Supporting Information: Above Room Temperature Ferroelectricity in Epitaxially Strained $KTaO_3$

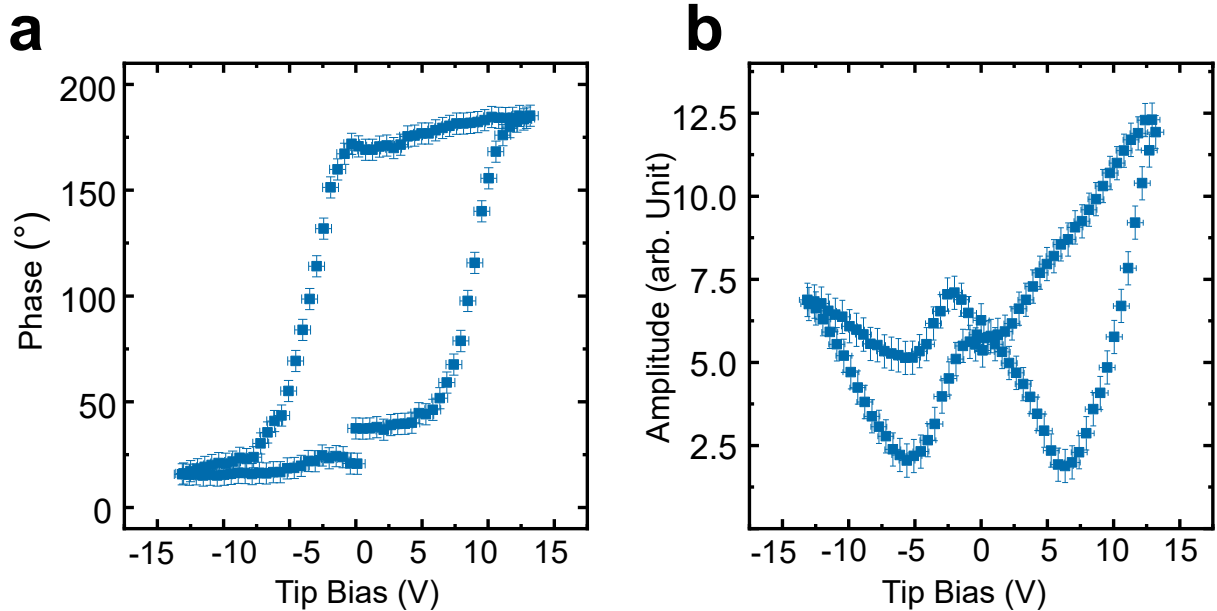

Fig. S12. **a** PFM switching using the $KTaO_3$/$SrRuO_3$/$DyScO_3$ sample showing a upward build-in field. **b** Magnitude of piezo response.

Supporting Information: Above Room Temperature Ferroelectricity in Epitaxially Strained KTaO$_3$


\end{document}